\documentclass[10pt]{iopart}

\usepackage{rotating,graphicx}

\usepackage{amsfonts}%
\usepackage{amssymb}%
\usepackage{caption}
\usepackage[utf8]{inputenc}
\usepackage[english]{babel} 
\usepackage[T1]{fontenc}
\usepackage{color}
\usepackage[colorlinks]{hyperref}
\usepackage{setspace}
\usepackage{layouts}
\usepackage{lipsum}  
\usepackage{bibentry}

\newcommand{\ignore}[1]{}
\newcommand{\nobibentry}[1]{{\let\nocite\ignore\bibentry{#1}}}

\nobibliography*

\begin{document}

\title[On the role of density fluctuations in the core turbulent transport of Wendelstein 7-X]{On the role of density fluctuations in the core turbulent transport of Wendelstein 7-X}
\author{D. Carralero$^1$, T. Estrada$^1$, E. Maragkoudakis$^1$, T. Windisch$^2$, J. A. Alonso$^1$, J.L. Velasco$^1$, O. Ford$^2$, M. Jakubowski$^2$, S. Lazerson$^2$, M. Beurskens$^2$, S. Bozhenkov$^2$, I. Calvo$^1$,  H. Damm$^2$, G. Fuchert$^2$, J. M. García-Regaña$^1$, U. Höfel$^2$, N. Marushchenko$^2$, N. Pablant$^3$, E. Sánchez$^1$, H. M. Smith$^2$, E. Pasch$^2$, T. Stange$^2$, and the Wendelstein 7-X team.}
\address{$^1$ Laboratorio Nacional de Fusión. CIEMAT, 28040 Madrid, Spain.}
\address{$^2$ Max-Planck-Institut für Plasmaphysik, D-17491 Greifswald, Germany.}
\address{$^3$ Princeton Plasma Phys Lab, 100 Stellarator Rd, Princeton, NJ 08540 USA} 
\ead{daniel.carralero@ciemat.es}

\begin{abstract}

A recent characterization of core turbulence carried out with a Doppler reflectometer in the optimized stellarator Wendelstein 7-X (W7-X) found that discharges achieving high ion temperatures at the core featured an ITG-like suppression of density fluctuations driven by a reduction of the gradient ratio $\eta_i = L_n/L_{T_i}$ [D. Carralero et al., Nucl. Fusion, 2021]. In order to confirm the role of ITG turbulence in this process, we set out to establish experimentally the relation between core density fluctuations, turbulent heat flux and global confinement. With this aim, we consider the scenarios found in the previous work and carry out power balance analysis for a number of representative ones, including some featuring high ion temperature. As well, we evaluate the global energy confinement time and discuss it in the context of the ISS04 inter-stellarator scaling. We find that, when turbulence is suppressed as a result of a reduction of $\eta_i$, there is a reduction of ion turbulent transport, and global performance is improved as a result. This is consistent with ITG turbulence limiting the ion temperature at the core of W7-X. In contrast, when turbulence is reduced following a decrease in collisionality, no changes are observed in transport or confinement. This could be explained by ITG modes being combined with TEM turbulence when the later is destabilized at low collisionalities.
\end{abstract}

\maketitle

\ioptwocol

\section{Introduction}\label{intro}

Optimized stellarators are regarded as a potential pathway for fusion reactors in which the general advantages of the stellarator approach (such as stationary pulses, absence of disruptions, etc.) are conserved while neoclassical (NC) fluxes, that traditionally dominate transport in stellarators, are reduced to acceptable levels. The most advanced device of this kind is currently Wendelstein 7-X (W7-X), which started operation in 2016 \cite{Wolf17}, and has recently achieved 200 MJ discharges and complete detachment in the island divertor \cite{Pedersen19}. While the results of the first experimental campaigns confirm that NC transport in W7-X has been reduced with respect to non-optimized stellarators \cite{Beidler19}, comparison of NC predictions with experimental measurements of the total energy transport implies that, at least under a wide range of scenarios, turbulence accounts for a large fraction of the total transport, even in the core \cite{Bozhenkov19,Dinklage18}. As a result, plasma performance has been below expectations from simulations which assumed mostly NC transport \cite{Turkin11,Bozhenkov19}: ion temperature at the core, $T_{i,core}$ remains "clamped" to a value of $T_{i,core} \leq 1.7$ keV regardlesss of heating power or configuration \cite{Beurskens21} and energy confinement time, $\tau_E$, falls typically below the scaling \cite{Fuchert20} when compared with the stellarator ISS04 database \cite{Yamada05}. Fortunately, a number of scenarios have been reported during the divertor campaign in which these limitations were overcome. The most relevant of them is the so called "High Performance Regime" (HP) in which plasma density is rapidly increased in a ECRH discharge by means of a series of injected pellets \cite{Bozhenkov20}. In this scenario, turbulence is suppressed \cite{Stechow20, Estrada21}, and transport drops to NC levels, thus achieving high central temperatures $T_{i,core} \simeq T_{e,core} \simeq 3$ keV and improved confinement levels, $\tau_E/\tau_{ISS04} \simeq 1.4$. Core ion temperatures exceeding the clamping value have also been reported in other regimes involving the use of NBI \cite{Ford19,Ford21} or the injection of impurities in the plasma \cite{Lunsford21, Stechow21}. What all these scenarios have in common is the formation of a steep density peaking leading to the suppression of turbulence and improved performance. This kind of scenario is not unique to W7-X: the HP regime is reminiscent of the "optimized confinement"\cite{Kick99} found on its predecessor, W7-AS, and similar regimes have been reported from other stellarators such as LHD \cite{Yamada00} or Heliotron-E \cite{Ida96}. As well, the equivalent "pellet-enhanced performance" in tokamaks is almost universal and has been identified in almost every major machine (for some examples on this, see the references listed in the Introduction of \cite{Bozhenkov20} or \cite{Estrada21}).\\

A widely accepted explanation for these enhanced regimes is based on the destabilization mechanism of ITG turbulence \cite{Wolfe86}, which is often related to a threshold on the gradient ratio $\eta_i = L_n/L_{T_i}$. Therefore, the build-up of a strong density gradient would reduce the $\eta_i$ parameter, thus stabilizing the ITG mode, which is typically expected to dominate transport in the ion-scale. This effect has been recently described in W7-X in a recent work which carried out a systematic characterization of microturbulence in the core of the machine \cite{Carralero21}. In it, a database was created of ion-scale turbulence measurements carried out with Doppler reflectometry (DR) in the standard configuration \cite{Klinger19}, including data representative of most relevant scenarios accessible to the DR during the last experimental campaign, as well as a number of examples in which high $T_{i,core}$ was achieved. The density fluctuation amplitudes were then compared with local measurements of gradients, finding that core density fluctuations drop substantially when $\eta_i$ drops below a certain value. This $\eta_i$-driven suppression of core turbulence was found in all the analyzed scenarios in which $T_{i,core}$ exceeded the clamping value, suggesting a link between the reduction of fluctions and the enhanced performance. However, this point remained mostly speculative as no attempt was made to relate the amplitude of density fluctuations and turbulent transport. In this work we aim to close that gap by analysing global turbulent transport studying its relation to fluctuation amplitude in the different regimes reported in previous work. The purpose of this is twofold: In the first place, to confirm the $\eta_i$ stabilization of ITG turbulence as the mechanism leading to the enhanced performance in W7-X. On a more general level, to investigate the relation (or lack thereof) between core microturbulence, turbulent transport and plasma global performance.  We ordered the remaining of this paper as follows: In Section \ref{Rev}, we review the main core turbulence regimes previously investigated. In Section \ref{Transport} we estimate global NC and turbulent transport values at the core for the different regimes and compare them to the fluctuation levels. Finally, we analyze the relation between core fluctution levels and global performance in Section \ref{global} and discuss the main implications of these results in Section \ref{end}. \\


\begin{figure*}
	\centering
	\includegraphics[width=\linewidth]{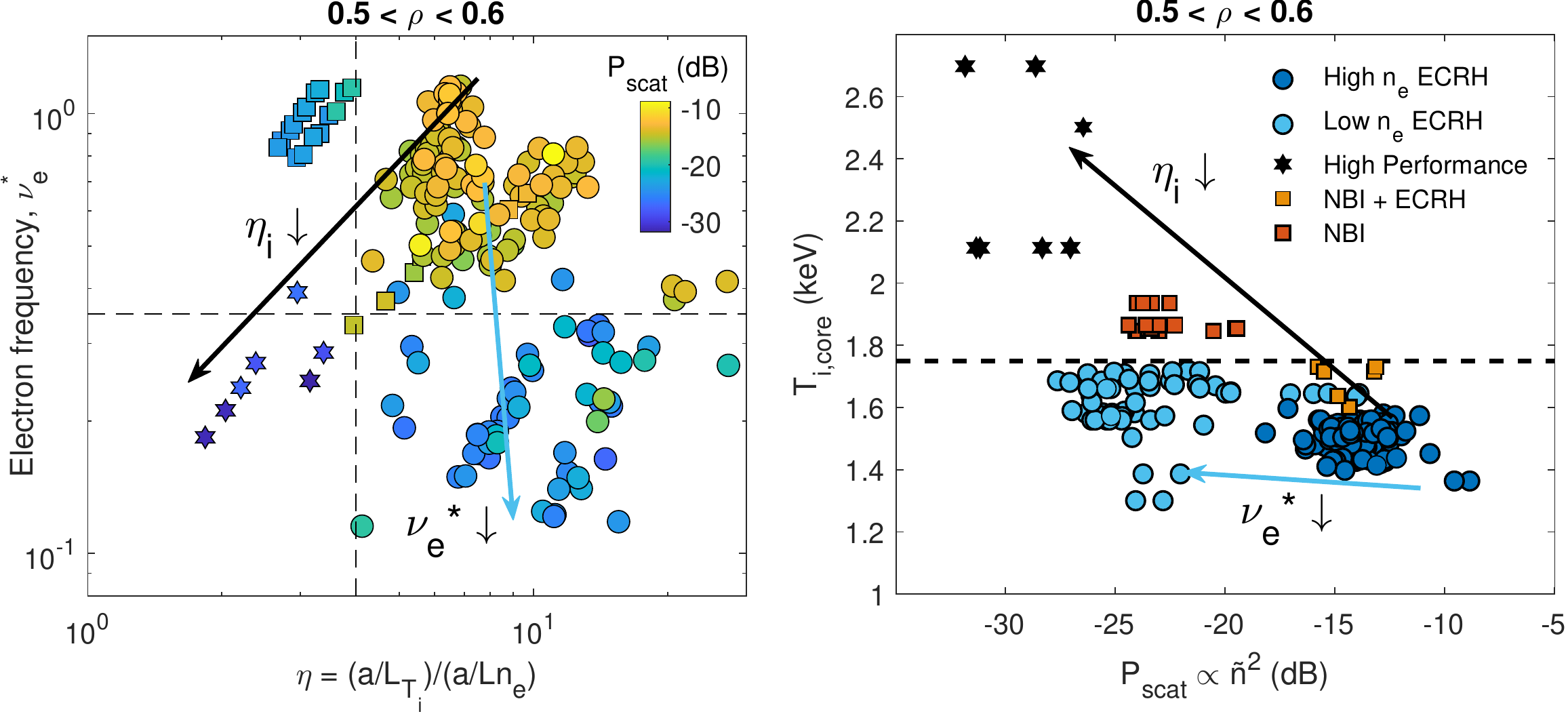}
	\caption{\textit{Summary of core turbulence regimes. Left) Measured scattered power, $P_{scat} \propto \delta n^2$, is represented by colorcode as a function of the temperature to density gradient ratio $\eta_i$ and the normalized electron collisionality $\nu_e^*$. Right) Core ion temperature is represented as a function of $P_{scat}$. Colors and symbols stand for the different regimes. In both cases, the high $a/L_n$/low collisionality pathways for reduced fluctuations are highlighted using black/blue arrows. These figures correspond respectively to Figure 14 and 15 in \cite{Carralero21}, where further details can be found.}}
	\label{fig0}
\end{figure*}

\section{Core fluctuation regimes}\label{Rev}

As already advanced in the Introduction, core density fluctuations have been recently characterized in the core of W7-X by means of a DR \cite{Estrada21,Carralero21}. As explained in detail in these references, this diagnostic launches a microwave beam into the plasma and  measures the power backscattered, $P_{scat}$, at the cut-off layer, which under certain approximations is roughly proportional to the second power of the density fluctuation amplitude, $P_{scat} \propto \tilde{n}^2$ \cite{Gusakov04,Blanco08}. To a good degree of approximation, this power backscattering typically happens at a certain, well known radial position, and is caused by density fluctuations of a certain wavenumber, $k_\perp$. This means that the flux surface and the wavenumber of the measured turbulence can be narrowly selected by setting the proper microwave beam wavenumber and incidence angle. In particular, these were set to probe the plasma core $0.5 < \rho < 0.75$ and to remain in the ion scale $k_\perp\rho_i \simeq 1$. Measurements were carried out with the V-band DR system installed in the AEA-21 port of W7-X (toroidal angle $\phi = 72^{\circ}$) \cite{Windisch15,Carralero20} with the fixed-angle antenna below the equator. This system overlooks the outer midplane of the elliptical section of the plasma, where curvature terms are strongest and therefore ITG modes are expected to be most unstable \cite{Banon20}. The characterization was carried out by constructing a database of discharges selected to cover most of the operational space accessible to the diagnostic, with density and heating power levels compatible with core fluctuation measurements. With this aim, three main groups of discharges were analyzed in the standard configuration: first, the bulk of the data consists of ECRH gas puff-fuelled shots, which were divided into low and high density subgroups ("LD ECRH" and "HD ECRH" in the following), following the discussion in \cite{Carralero21}. On top of these data, discharges from two scenarios with $T_{i,core}$ exceeding the clamping value were analyzed: first, a number of ECRH, pellet-fueled high performance discharges ("HP" in the following). Second, an NBI-improved scenario comprising a phase heated with both ECRH and NBI power ("ECRH+NBI" in the following) with $T_{i,core}$ still below the clamping, and a NBI-dominated phase ("NBI" in the following) in which ECRH heating was reduced to around $0.5$ MW and higher $T_{i,core}$ values were achieved. More details on the scenarios and examples of particular discharges can be found in \cite{Carralero21}. In total, the database consisted of over 150 points from 18 different discharges.\\

The main results from the turbulence characterization are summarized in Figure \ref{fig0}, in which discharges from the different scenarios are represented using different symbols. On the left plot, discharges are represented as a function of $\eta_i$ and electron normalized collisionality $\nu_e^*$ using a color code to indicate the amplitude of fluctuations (measured in the $0.5 < \rho < 0.6$ radial region using the $P_{scat}$ proxy). As can be seen, points can be classified in three main groups: discharges featuring both high $\eta_i$ and collisionality values display strong density fluctuations. Then, fluctuations seem to be suppressed below two threshold values: one for the gradient ratio and another for the collisionality, with critical values around $\eta_i \simeq 4$ and $\nu_e^* \simeq 1/3$ respectively. On the right plot of Figure \ref{fig0}, the evolution of $T_{i,core}$ with fluctuations is displayed for the same scenarios. In it, two different trends can be seen: in the cases when fluctuations are reduced along with a low $\eta_i$ value -the "$\eta_i$ pathway"- such reduction leads to an increase of $T_{i,core}$, which can go above the clamping value (indicated in the figure as a dashed line). Instead, in the LD ECRH case $\eta_i$ remains unchanged and fluctuations are reduced along the collisionality -the "$\nu_e^*$ pathway". In this second case, $T_{i,core}$ does not improve significantly and remains below the clamping value. These results were interpreted in \cite{Carralero21} as follows: HD ECRH and NBI+ECRH discharges would fall in the "baseline" scenario, in which ITG turbulence would dominate transport due to the high $\eta_i$ values, leading to the observed high $P_{scat}$, degrading confinement and limiting the value of $T_{i,core}$. When $\eta_i$ is reduced below the threshold -as in NBI and HP discharges-, ITG turbulence would be stabilized and this would explain the reduced fluctuations and improved performance. In these two cases, the high values of $\nu_e^*$ would keep TEM turbulence stable. However, when $\nu_e^* < 1$ values are achieved, these modes could be destabilized (see eg. \cite{Ryter05} for the description of a similar effect in tokamaks). Since TEMs are destabilized in a different toroidal region than ITG modes (namely, in the triangular section, where trapped particle population is higher), one possible outcome of this would be that due to the interaction of TEM and ITG modes, the strongest fluctuations would no longer be found at the elliptical section, thus moving away from the region probed by the DR while not necessarily affecting global transport. This would be consistent with the observed drop in $P_{scat}$  while $T_{i,core}$ still seems to be bounded by the clamping. However, as already pointed out in the Introduction, since the DR can only measure local density fluctuations and not heat fluxes, this interpretation remained speculative. In order to support it, it would be required to show how the reduction of fluctuations along the $\eta_i$ pathway corresponds to a reduction of turbulent transport in the same flux surfaces. Similarly, if the TEM destabilization hypothesis is correct, the reduction of fluctuations along the $\nu_e^*$ pathway should lead to no substantial changes of transport instead.

\section{Evaluation of core heat transport}\label{Transport}

\begin{figure*}
	\centering
	\includegraphics[width=\linewidth]{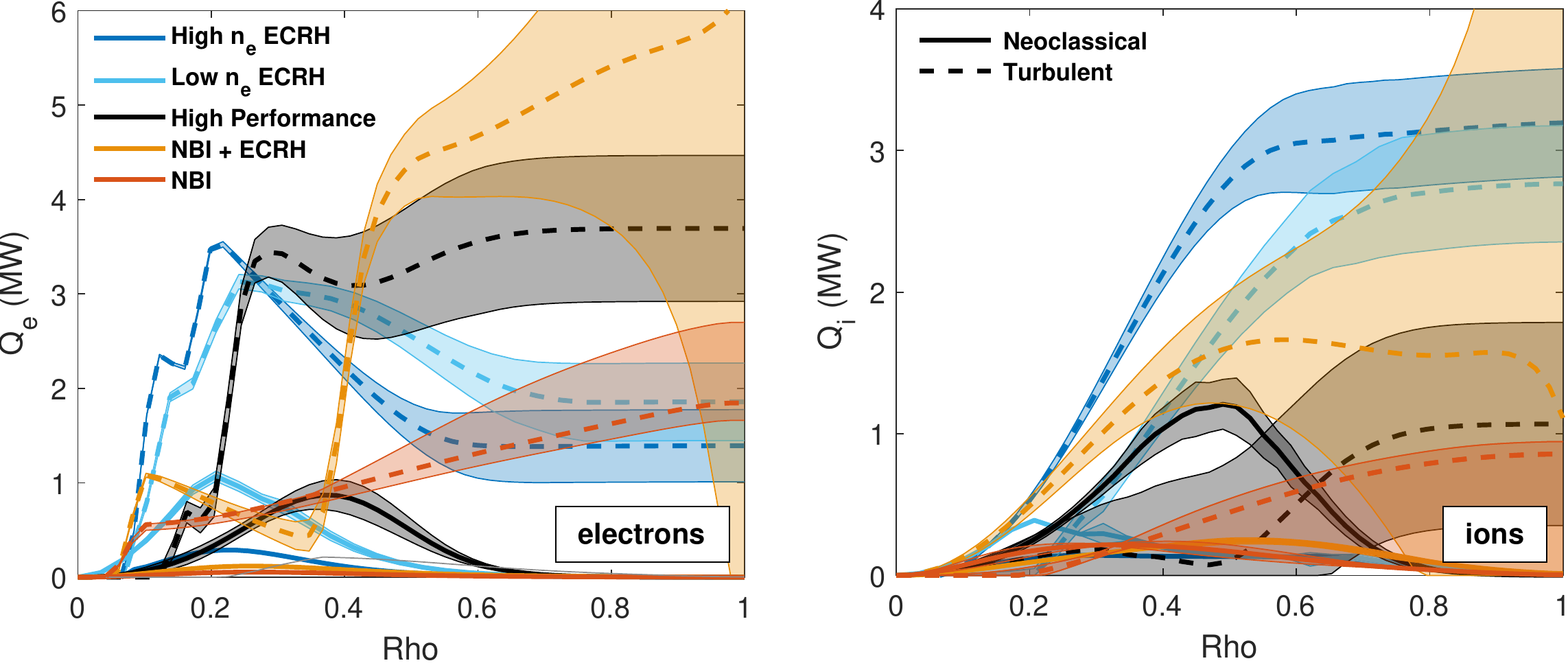}
	
	\caption{\textit{Heat flux profile comparison for electrons (left) and ions (right). As in Figure \ref{fig0}, color indicates scenario. Dashed/solid lines stand for.  turbulent/neoclassical flux.}}
	\label{fig2a}
\end{figure*}
In order to evaluate local turbulent transport, we resort to a power balance analysis. This technique, already applied in W7-X \cite{Bozhenkov20} and other stellarators \cite{Stroth98,Dinklage13}, provides the flux surface-averaged heat flux, $Q_\alpha$ for a given species $\alpha$ by assuming a stationary state and integrating source and sink terms in the energy conservation equation inside said flux surface. This total heat flux is considered to be the combination of neoclassical and turbulent transport. Neoclassical transport, $Q^{NC}$, can be calculated from plasma density and temperature profiles using numerical models \cite{Beidler11}, so turbulent transport can be estimated as the remaining part of total transport, $Q^{turb}_\alpha = Q_\alpha-Q^{NC}_\alpha$.\\

\begin{figure*}
	\centering
	\includegraphics[width=\linewidth]{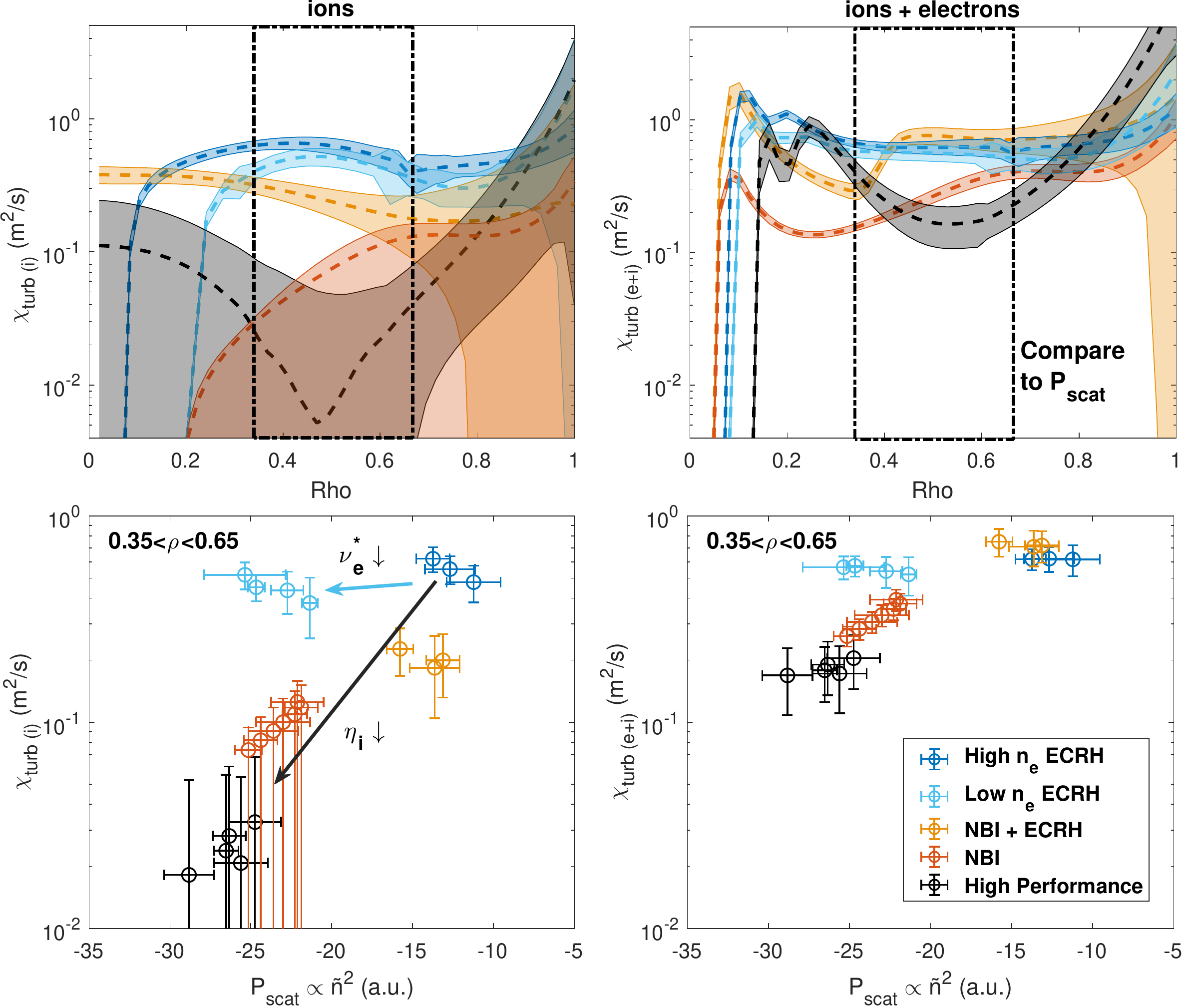}
	
	\caption{\textit{Transport coefficients. Top) Ionic and total profiles of turbulent transport coefficient $\chi_{turb}$. Bottom) $\chi_{turb}$ is represented as a function of the local amplitude of fluctuations measured by the DR.}}
	\label{fig2b}
\end{figure*}

 In this analysis, we take a subset of the discharges included in the DR measurement database presented in \cite{Carralero21}, including examples of the previously introduced HD ECRH ($\#180920017$), LD ECRH ($\#180920013$), HP ($\#180918041$), NBI+ECRH ($\#180919039, t \simeq 3.5 s$) and NBI regimes  ($\#180919039,$ $t \simeq 4.5 s$). For each of them, we calculate neoclassical transport using both Neotransp code (using tabulated mono-energetic coefficients from DKES \cite{Rij89}) and KNOSOS code \cite{Velasco19, Velasco21} (which recalculates coefficients in order to take properly into account the tangential magnetic drift in the low collisionality cases \cite{Calvo17}), both of which provide similar results in the analysed cases. Then, energy source terms are calculated for each of the heating systems: In the case of ECRH, Travis code \cite{Maru14} is used to determine the microwave power deposition profile. In the case of NBI, this is calculated with the BEAMS3D code \cite{Lazerson21,Lazerson21b}. Once these source terms are known, a simplified power balance can be calculated at the flux-surface simply by integrating them along with the electron-ion collisional thermalization term, which is typically the only remaining sink/source relevant at the plasma core. This simplified analysis has its limitations, however: in the first place, there are several sink terms which may become relevant as the edge is approached (such as radiation losses in electrons or CX in ions). As well, the reliability of the calculated $Q^{NC}$ values depends on the quality of the density and temperature gradient measurements used as an input, which tends to deteriorate close to the plasma center. Furthermore, while the stationary plasma hypothesis is typically appropriated, in some cases (namely, the HP scenario) the evolution of profiles is significantly faster than typical energy transport times, and may thus play a relevant role. As a result, the results of this simplified calculation are only valid for a limited radial range, which will limit the scope of the discussion. A conservative estimation, based in more detailed analysis of the topic presented in \cite{Bozhenkov20}, would yield a radial valid range of $0.2 < \rho < 0.7$. Finally, a word of caution is in order regarding the NBI scenario: in this case, $T_i \simeq T_e$ for most of the profile. However, differences of the order of tens of eV are within the error bars of the diagnostics which, given the strong collisionality of this high density scenario, would lead to unreasonably high interspecies heat transfer, particularly for the outer mid-radius. Following the discussion in \cite{Ford21}, where this problem is examined in greater detail, $Q_i$ is assumed to be properly bounded by assuming $Q_i^{NC} < Q_i < Q_i^{t}$, where $Q_i^t$ is the heat flux obtained assuming thermalization and no inter-species heat exchange. These bounds substitute diagnostic errorbars (less conservative) for the NBI scenario in Figures \ref{fig2a} and \ref{fig2b}.\\

The results of the analysis are represented in Figure \ref{fig2a}, where $Q^{NC}$ and $Q^{turb}$ are represented for both electrons and ions in each of the discussed scenarios. As can be seen, electron transport is dominated by the turbulent contribution for the whole radial profile, with the neoclassical contribution well below it in all cases. Interestingly, the NBI case features low $Q_e$ values at the core (as only $0.5$ MW of ECRH power is injected), followed by a monotonous increase up to the separatrix (NBI power is deposited on both species across the whole radius): Since ions and electrons are likely close to thermalization -as discussed before- the large reduction of $Q_e$ as a result of the transfer to the ions in the HD/LD ECRH or NBI+ECRH scenarios is not observed in this case. Regarding the ions, a similar picture is observed in the valid radial range for the ECRH and NBI+ECRH scenarios. Similarly, the NBI case displays a much slower increase of $Q_i$, reflecting the reduced transfer from the electrons. Turbulent transport dominates over neoclassical at an outer radial position than in the ECRH and NBI+ECRH scenarios ($0.35\leq \rho \leq 0.65$), although a precise determination is not possible due to the large errorbars. Finally, turbulent transport is strongly suppressed in the high performance case \cite{Bozhenkov20}, which features NC levels of $Q_i$ up to $\rho \simeq 0.6$. \\

In order to better quantify the importance of turbulent transport, these  profiles are used to calculate the flux-surface averaged turbulent heat transport coefficient, 

\begin{equation}
	\chi_{turb,\alpha} = -\frac{2}{3}\frac{Q^{turb}_\alpha}{n_\alpha\partial_r T_\alpha}.
\end{equation}

This can then be compared to the local DR measurements of scattered power in the radial region $0.35 < \rho < 0.65$, which can be taken as a proxy for density fluctuations as $P_{scat} \propto \tilde{n}^2$, as already explained. The results are displayed in Figure \ref{fig2b}, where the relevant region is highlighted by the dashed-dotted black line: As can be seen, the ion transport coefficient roughly mirrors the already described trends for $Q_i^{turb}$: NBI and HP scenarios display substantially reduced coefficients. When both ions and electrons are considered together, a similar picture is obtained, although the differences are strongly smoothed. This is consistent with the results displayed in Figure \ref{fig2a} and indicates that the main differences in transport between scenarios are related to the ion channel.  When the resulting coefficients are compared to fluctuations, two different trends can be observed in the data: when HD ECRH, NBI+ECRH, NBI and HP scenarios are considered, the reduction of fluctuations –which was linked to a reduction of $\eta_i$ and to enhanced performance in the previous study- leads to a clear reduction of the ion turbulent transport coefficient. When the transport of both species is considered the same trend is observed, although the effect is considerably weaker, indicating again that this effect does not affect electrons (or at least, affects them considerably less). Instead, when the evolution between HD and LD ECRH discharges is considered, the decrease of turbulence amplitude is not related to a variation of the flux surface-averaged turbulent transport in either species. A final note regarding the NBI scenario is in order: Given the previously discussed problem with error estimation for $Q_i$ in these shots, there is no lower bound for $Q_{i,turb}$. As a result, while the central values of the observations suggest that $\chi_{turb(i)}$ is higher in the NBI scenario than in the HP one, this can not be stated with certainty within the error bars. In any case, this is unlikely to affect the qualitative trend relating $\chi_{turb(i)}$ to $P_{scat}$, which seems to hold in a any case for $\chi_{turb(e+i)}$ (total transport does not need to take into account inter-species heat transfer and is therefore properly bounded also in the NBI scenario).

\section{Impact on global performance}\label{global}
  
\begin{figure}
	\centering
	\includegraphics[width=0.8\linewidth]{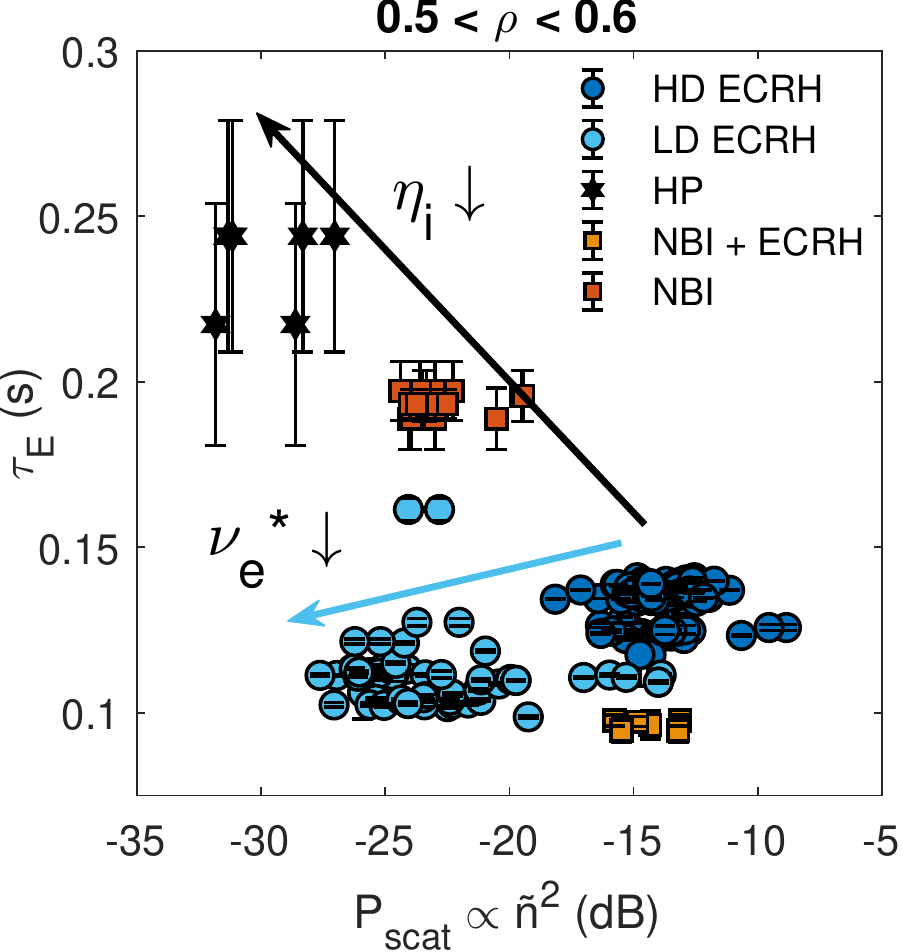}
	
	\caption{\textit{Global performance as a function of measured fluctuation amplitude. Energy confinement time, $\tau_E$, is represented o as a function of core density fluctuations. Colors and symbols stand for the different scenarios, as in Figure \ref{fig0}.}}
	\label{fig3a}
\end{figure}

Once the relation between local fluctuations and transport has been established at the flux-surface level, this analysis can be complemented by evaluating the impact of core fluctuations on global confinement. For this, the density fluctuation proxy $P_{scat}$ measured at the  previously discussed $0.5 < \rho < 0.6$ radial region is compared to the global energy confinement time, $\tau_E$, calculated over the duration of the DR frequency ramp. While most of the data points are stationary over this interval, that is not always the case: as previously discussed, HP discharges undergo fast changes in profiles (and therefore in the stored energy, $W_{dia}$) over this period, thus yielding large error bars. In this sense, the HP points must be regarded as a conservative estimation of $\tau_E$, with values potentially lower than those achieved at the time of maximum $W_{dia}$. Since the detailed power balance calculation required in the previous section to evaluate $\chi$ is no longer necessary, all the discharges in the data base could be included for this. The results, displayed in Figure \ref{fig3a}, are consistent with the picture provided in the previous section: when lower fluctuations are achieved reducing $\eta_i$ and thus core turbulent transport, global confinement is improved. Instead, when fluctuations drop following a reduction in collisionality, neither turbulent transport nor global confinement are improved. This observation is also consistent with the relation between $T_{i,core}$ -which was advanced in \cite{Carralero21} as a proxy for global confinement- and the amplitude of core turbulence, as seen in Figure \ref{fig0}. However, Figure \ref{fig3a} presents an overly simplified picture, as it is potentially ignoring general dependencies of confinement on other plasma parameters, such as heating power or density. For example a degradation in confinement is observed for the LD ECRH and NBI+ECRH scenarios with respect to the HD ECRH, all of which displayed similar transport coefficients in Figure \ref{fig2b}, but feature rather different densities.\\

\begin{figure*}
	\centering
	\includegraphics[width=\linewidth]{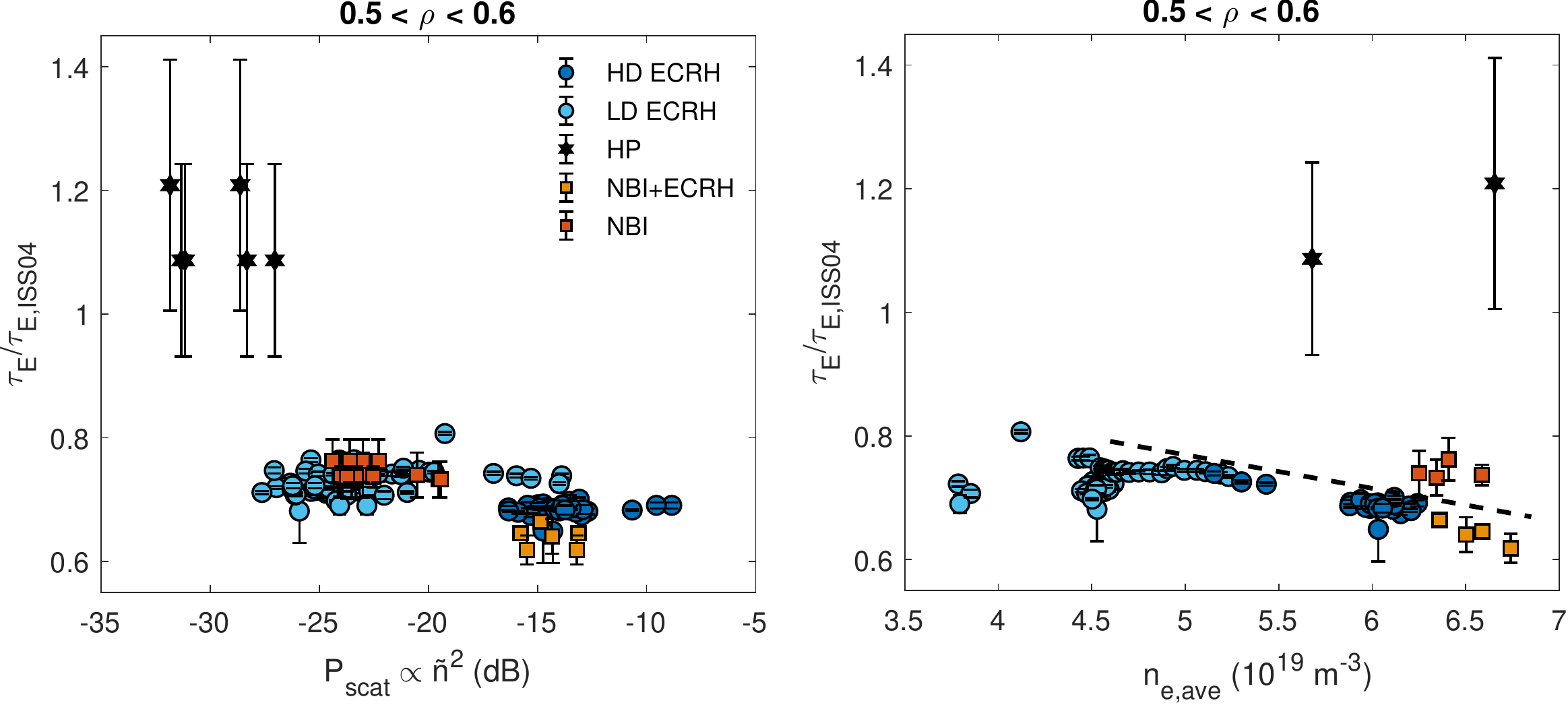}
	
	\caption{\textit{Evaluation of normalized global performance. Confinement ratio with respect to the ISS04 scaling is represented respectively as a function of the core fluctuation values and line averaged plasma density on the left/right plot. Colors and symbols stand for the different scenarios, as in Figure \ref{fig0}. On the right plot, the confinement degradation typically associated to density (see \cite{Fuchert20}) is highlighted as a dashed line.}}
	\label{fig3b}
\end{figure*}

In order to disentangle the impact of turbulence with other parameters potentially affecting $\tau_E$, it has been normalized using the ISS04 stellarator scaling, which for a given W7-X configuration can be stated as

\begin{equation}
\tau_{E,ISS04} \propto \bar{n}_e^{0.54}P_h^{-0.61},
\end{equation}

where $\bar{n}$ is the line averaged density and $P_h$ is the heating power. This is represented on the left plot of Figure \ref{fig3b}: in it, the usual clear improvement of the HP scenario is conserved and the normalized confinement of the ECRH phases decreases now with fluctuation amplitude, which suggests that the reduction of confinement of the LD ECRH scenario in Figure \ref{fig3a} is probably explained by the density dependency of $\tau_E$. However, LD ECRH and NBI scenarios now seem to increase similarly with the reduction of fluctuations, while the NBI+ECRH scenario still stays below the HD ECRH for similar fluctuations. This apparent inconsistency can be explained by taking into account the previously mentioned work on the subject \cite{Fuchert20}, which has shown that typical values of $\tau_E/\tau_{E,ISS04} \simeq 0.8$ can be found for low and intermediate densities, with the ratio degrading down to $\tau_E/\tau_{E,ISS04} \simeq 0.6-0.7$ for high plasma density, depending on the level of radiation. Indeed, this picture is recovered when $\tau_E/\tau_{E,ISS04}$ is represented as a function of density, as in the right plot of Figure \ref{fig3b}: in it, all gas-puffed scenarios using ECRH align roughly as in the mentioned reference (the trend is marked in a dashed line). However, the NBI and particularly the HP scenarios stand out of that trend achieving higher $\tau_E/\tau_{E,ISS04}$ values despite their high density. Once again, when the $\eta_i$ pathway is followed, the reduction of fluctuations leads to a clear improvement in global confinement with respect to the usual scaling. Instead, when it is achieved following the $\nu_e^*$ pathway, performance seems to remain unaffected.\\

\section {Conclusions}\label{end}

In summary, we have been able to establish a clear link between core microturbulence amplitude, turbulent transport and global performance in the optimized stellarator W7-X. In particular, in discharges with high $T_{i,core}$ values, the formation of strong density gradients leads to the reduction of the ITG-driving gradient ratio $\eta_i$, causing a clear suppression of density fluctuations in the $k_\perp\rho_i \simeq 1$ ion scale. In this case, a strong reduction of the ion heat transport coefficient, $\chi^{turb}_i$, is observed and the global confinement rises over the values typically reported for similar densities. All these findings are consistent with the hypothesis of ITG turbulence limiting $T_{i,core}$ and thus performance. On the other hand, it is also found that, while such reduction in core fluctuations seems to be a necessary condition for high performance, it is not a sufficient one: this is exemplified by the LD ECRH scenario, in which the drop in fluctuations is associated to a drop in $\nu_e^*$ rather than to $\eta_i$, and has thus no effect in $\chi^{turb}_i$ nor in confinement equivalent to the turbulence reduction. This different global effect is consistent with the proposed hypothesis of a change in the dominant turbulent modes that would now include a contribution from TEMs, which would be destabilized by the low collisionality. In this case, the observed reduction in fluctuations would be mostly instrumental (as the unstable region would move away from the DR beam) and thus not necessarily affect transport. Whether this combination of TEM and ITG modes could drive turbulent transport without significantly affecting $\chi_{turb}$ or $\tau_E$, remains an open question and will be addressed in future work. Beyond W7-X, by this result is of great importance for the future understanding of the fundamental physics of turbulence, as it indicates how global gyrokinetic codes being used to describe turbulent transport can be properly validated by local measurements of fluctuations.\\

\section*{Acknowledgments}

The authors acknowledge the entire W7-X team for their support. This work has been partially funded by the Spanish Ministry of Science and Innovation under contract number FIS2017-88892-P and partially supported by grant ENE2015-70142-P, Ministerio de Economía y Competitividad, Spain and by grant PGC2018-095307-B-I00, Ministerio de Ciencia, Innovación y Universidades, Spain. This work has been sponsored in part by the Comunidad de Madrid under projects 2017-T1/AMB-5625 and Y2018/NMT [PROMETEO-CM]. This work has been carried out within the framework of the EUROfusion Consortium and has received funding from the Euratom research and training programme 2014-2018 and 2019-2020 under grant agreement No 633053. The views and opinions expressed herein do not necessarily reflect those of the European Commission.\\


\begin{thebibliography}{99}
	
	

\bibitem{Carralero21} D. Carralero, T. Estrada, E. Maragkoudakis, Nucl. Fusion 61 096015 (2021).
\bibitem{Wolf17} R. C. Wolf, A. Ali, A. Alonso et al. Nucl. Fusion {\bf 57}, 102020 (2017). 
\bibitem{Pedersen19} T. Sunn Pedersen, R. König, M. Jakubowski et al., Nucl. Fusion 59 096014 (2019).
\bibitem{Beidler19} C. D. Beidler, H. M. Smith, A. Alonso, et al., Nature 596, pages 221–226 (2021)
\bibitem{Bozhenkov19} S. Bozhenkov on behalf of the W 7-X team, 46th EPS Conference on Plasma Physics, Milan (2019).
\bibitem{Dinklage18} Dinklage, A., Beidler, C.D., Helander, P. et al. Nature Phys 14, 855–860 (2018).
\bibitem{Turkin11} Y. Turkin, C.D. Beidler, H. Maaberg et al., Phys. Plasmas 18 022505 (2011).
\bibitem{Beurskens21} M.N.A. Beurskens, S.A. Bozhenkov, O. Ford et al., accepted for publication in Nucl. Fusion (2021).
\bibitem{Fuchert20} G. Fuchert, K.J. Brunner, K. Rahbarnia et al., Nucl. Fusion \textbf{60} 036020, (2020). 
\bibitem{Yamada05} H. Yamada, J.H. Harris, A. Dinklage et al., Nucl. Fusion 45 1684 (2005).
\bibitem{Bozhenkov20} S.A. Bozhenkov, Y. Kazakov, O.P. Ford et al., Nucl. Fusion \textbf{60} 066011, (2020).
\bibitem{Stechow20} A. v.  Stechow, 22nd International Stellarator and Heliotron Workshop, Madison, Wisconsin, USA (2019).
\bibitem{Estrada21} T. Estrada, D. Carralero, T. Windisch et al., Nucl. Fusion  {\bf 61} 046008 (2021).
\bibitem{Ford19} O. P. Ford, S.Bozhenkov, M.Beurskens et al., 46th EPS Conference on Plasma Physics, Milan (2019).
\bibitem{Ford21} O. Ford, M. Beurskens, S. Bozhenkov et al., submitted to Plasma Phys. Control. Fusion (2021)
\bibitem{Lunsford21} R. Lunsford, C. Killer, A. Nagy et al., Phys. Plasmas 28, 082506 (2021)
\bibitem{Stechow21} A. von Stechow, Z. Huang, J.-P. Bähner et al., 47th EPS Conference on Plasma Physics, Sitges (2021).
\bibitem{Kick99} M. Kick, H. Maaßberg, M. Anton et al.,  Plasma Phys. Control. Fusion 41 A549 (1999)
\bibitem{Yamada00} H.Yamada, R.Sakamoto, S.Sakakibara et al., 27th EPS Conference on Contr. Fusion and Plasma Phys. Budapest, 12-16 June 2000
\bibitem{Ida96} K. Ida, K. Kondo, K. Nagasaki et al., Phys. Rev. Lett {\bf 76} 8 (1996).
\bibitem{Wolfe86} S.M. Wolfe, M. Greenwald, R. Gandy et al., Nucl. Fusion 26 329 (1986)
\bibitem{Klinger19} Klinger T., Andreeva T. and Bozhenkov S. et al. Nucl. Fusion 59 112004 (2019).
\bibitem{Gusakov04} E. Z. Gusakov, and A. V. Surkov,  Plasma Phys. Control. Fusion {\bf 46} 1143 (2004).
\bibitem{Blanco08} E. Blanco and T. Estrada, Plasma Phys. Control. Fusion {\bf 50}  095011 (2008).
\bibitem{Windisch15} T. Windisch, T. Estrada, M. Hirsch et al., 42nd EPS Conference on Plasma Physics, Lisbon (2015).
\bibitem{Carralero20} D. Carralero, T. Estrada, T. Windisch et al., Nucl. Fusion \textbf{60} 106019 (2020).
\bibitem{Banon20}  A. Bañón Navarro, G. Merlo, G. G. Plunk et al., Plasma Phys. Control. Fusion {\bf 62} 105005 (2020).
\bibitem{Ryter05} F. Ryter, C. Angioni, A. G. Peeters et al., Phys. Rev. Lett. \textbf{95}, 085001 (2005).
\bibitem{Stroth98} U. Stroth, J. Baldzuhn, J. Geiger et al., Plasma Phys. Control. Fusion 40 1551 (1998).
\bibitem{Dinklage13} A. Dinklage, M. Yokoyama, K. Tanaka et al., Nucl. Fusion 53, 063022 (2013).
\bibitem{Beidler11} C.D. Beidler, K. Allmaier, M.Yu. Isaev et al., Nucl. Fusion 51 076001 (2011).
\bibitem{Rij89} van Rij and Hirshman, Phys FLuids B 1, 563 (1989).
\bibitem{Velasco19} Velasco J.L., Calvo I., Parra F. and García-Regaña J.M. 2019 J. Comp. Phys. 109512.
\bibitem{Velasco21} J. L. Velasco, I. Calvo, F. I. Parra et al., accepted for publication in Nucl. Fusion (2021), https://arxiv.org/abs/2106.01727.
\bibitem{Calvo17} Calvo I., Parra F.I. and Velasco J.L. et al 2017 Plasma Phys. Control. Fusion 59 055014.
\bibitem{Maru14} Marushchenko N.B. et al 2014 Comput. Phys. Commun 185 165.
\bibitem{Lazerson21} Samuel A. Lazerson, Oliver P. Ford, Carolin Nuehrenberg et al., Nucl. Fusion 60 076020 (2020).
\bibitem{Lazerson21b} Samuel A. Lazerson, David Pfefferlé, Michael Drevlak et al., Nucl. Fusion 61 096005 (2021).




















%

	
\end{thebibliography}
\end{document}